\documentclass[
               prd,nofootinbib,superscriptaddress,
               showpacs,preprintnumbers,
               amsmath,amssymb]
              {revtex4}
\usepackage{graphicx,color}
\usepackage{epsfig}

\newcommand{\loe}{\lesssim}
\def\be{\begin{equation}}
\def\ee{\end{equation}}
\newcommand{\bea}{\begin{eqnarray}}
\newcommand{\eea}{\end{eqnarray}}

\providecommand{\href}[2]{#2}

\begin{document}

\preprint{INR/TH-37-2004}
\preprint{BUHEP-04-10}

\title{Holes in the ghost condensate}

\author{D. Krotov}
\affiliation{Institute for Nuclear Research
of the Russian Academy of Sciences,\\
  60th October Anniversary prospect 7a, Moscow 117312, Russia}
\affiliation{Moscow State University, Department of Physics,\\
  Vorobjevy Gory, Moscow, 119899, Russia}
\affiliation{Institute of Theoretical and Experimental Physics,\\
  B. Cheremushkinskaya, 25, Moscow, 117259, Russia}
\author{C.~Rebbi}
\affiliation{Department of Physics---Boston University\\
  590 Commonwealth Avenue, Boston MA 02215, USA}
\author{V.~Rubakov}
\affiliation{Institute for
Nuclear Research of the Russian Academy of Sciences,\\
  60th October Anniversary prospect 7a, Moscow 117312, Russia}
\author{V.~Zakharov}
\affiliation{Max-Planck Institut f\"ur Physik\\
F\"oringer Ring 6, 80805, M\"unchen, Germany}

\begin{abstract}

In a recently proposed model of ``ghost condensation'', spatially
homogeneous states
may mix, via
tunneling, with inhomogeneous states which are
somewhat similar to bubbles in the
theory of false vacuum decay, the corresponding
bubble nucleation rate being
exponentially sensitive to the ultraviolet completion of the
model. The conservation of energy and charge requires that the energy
density is negative and the field is strongly unstable in a part of
the nucleated bubble. Unlike in the theory of false vacuum decay, this
region does not expand during subsequent real-time evolution. In the
outer part, positive energy outgoing waves develop, which eventually
form shocks. Behind the outgoing waves and away from the bubble
center, the background settles down to its original value. The outcome
of the entire process is thus a microscopic region of negative energy
and strong field ---  ``hole in the ghost condensate'' --- plus a
collection of outgoing waves (particles of the ghost condensate field) 
carrying away finite energy.

\end{abstract}

\pacs{11.10.Lm,11.90.+t}
\maketitle

\section{Introduction and summary}
\label{sec:intro}

In view of the evidence for the accelerated expansion of the Universe,
several attempts have been made recently to construct models in which gravity
is modified at large
distances~\cite{Charmousis:1999rg,
Kogan:1999wc,DGP,Jacobson:2001yj,
Freese:2002sq,Carroll:2003wy,AH,Holdom:2004yx}.
One approach, dubbed ghost condensation~\cite{AH}, invokes a scalar field
with   unconventional kinetic term, like in models of
k-essence~\cite{Armendariz-Picon:2000dh} but with the action depending on the
derivatives only,
\be
    S = M^4 \int~d^4x~ P(X)
\label{1**}
\ee
where
\[
  X = \partial^\mu \phi \partial_\mu \phi
\]
(space-time signature $(+,-,-,-)$). One views this model as an
effective theory valid at energies  below some cutoff.
Assuming that $P^{\prime} (0) < 0$, excitations about a state $\phi =0$ are
ghosts, hence the name of the model. A proptotype example of the
``potential'' having this property is
\be
 P(X) = \frac{1}{2}(X - 1 )^2
\label{P}
\ee
When gravity is switched off, the scalar
theory is perturbatively stable provided that the background has
the Lorentz-violating form,
\be
    \phi = ct
\label{1*}
\ee
where the constant $c$ is such that
\begin{eqnarray}
    P^{\prime}(c^2) &\geq& 0
\label{1+} \\
    2 P^{\prime \prime}(c^2) c^2 + P^{\prime}(c^2) &\geq& 0
\label{1++}
\end{eqnarray}
(for the potential (\ref{P}) these inequalities imply $c \geq 1$).
Cosmological evolution drives the field to a special point 
\be
 c= c_*
\label{*!}
\ee
where
\[
     P^{\prime}(c_*^2) = 0
\]
($c_* = 1$ for the potential (\ref{P})).
At the level of small perturbations about this background, 
the model has
interesting phenomenology, as discussed in
Refs.~\cite{AH,Dubovsky:2004qe,Peloso:2004ut}.

There are several points to discuss beyond the perturbation theory. One is
the behavior of the ghost condensate near sources of strong 
gravitational field,
e.g., near black holes~\cite{Frolov:2004vm}. In this paper we address another
issue, namely, quantum stability of the ghost condensate. As we discuss in
Section~2, even in the absence of gravitational interactions, there exist
inhomogeneous configurations of the scalar field, bubbles,
with which the state
(\ref{1*}) can mix via tunneling. 
We will see that
the tunneling exponent is
dominated by the ultraviolet (UV) properties of the theory, so the decay rate
is exponentially sensitive to the UV completion of the model, in accord with
Ref.~\cite{AH}. If the UV cutoff is well below the scale $M$, the bubble
nucleation rate is small.

Still, it is of interest to understand the further evolution
 of the bubbles which can
nucleate via tunneling. In conventional scalar theories with false vacua
(local minima of the scalar potential), a bubble of the true vacuum, once
created, expands practically with the speed of
light~\cite{Kobzarev:1974cp,Coleman:1977py}, and the true vacuum
eventually occupies the entire space. 
The difference between the energies of the false and true vacua
is released into the bubble wall, whose energy thus tends to infinity
 at asymptotically large time.
We will see in Section~3 that in the ghost
condensation model the situation is quite different. 
In the central part of a bubble, the energy density is negative and,
furthermore, the condition (\ref{1+}) (and also (\ref{1++}))
is violated. The system is unstable there, so the ultimate fate of the
field in this region cannot be understood in the UV-incomplete theory.
The point, however, is that this part does not expand and remains of
microscopic size at all times; this is a ``hole in the ghost condensate''.
Moreover, there is no energy flow through the boundary of this
hole\footnote{This is true, strictly speaking, only at the level of
classical field theory with the action (\ref{1**}). 
Nevertheless, under an assumption
that the energy density is bounded from below, the energy release from
the hole is finite, so a possible effect of the hole on the outer part
is minor in the quantum theory as well.}, so the hole has little,
if any, effect on
the outer part of the bubble. 

In the outer part, a shock wave (or series of shock waves)
is formed\footnote{Shock
  waves/kinks emerge in many models whose Lagrangians are non-linear in
  derivatives, see, e.g., Refs.~\cite{Felder:2002sv,Gibbons:2003yj}.}
which propagates
outwards. At the level 
of classical field theory with the action (\ref{1**}),
the formation of a shock wave means a singularity after which the solution
ceases to exist. Resolving this singularity would be possible in UV complete
theory only. 
Assuming that the singularity is smoothened out by the UV effects, we
study the entire evolution of the bubble. We find that the ghost
condensate behind the shock wave settles down to its original value,
$\dot{\phi} =c$, $\vec{\nabla} \phi =0$, so the energy of the shock wave
does not increase in time. Therefore, 
the amplitude of the shock wave decreases in time;
in the quantum theory the wave ultimately
decays into quanta of the ``ghost field'' $\phi$.
In the case of the special background (\ref{*!}) all these properties
follow from the energy conservation, while for the general initial
background 
$\dot{\phi} = c > c_*$ we have found them by numerical
simulations\footnote{For general initial background, energy
conservation does not forbid that the background behind 
the shock wave has lower energy density than the
original background in front of the shock wave. Were this the case,
the energy of the shock wave would increase in time, while the new
background would eventually occupy the entire space outside the
hole.
The whole process
would then be analogous to the false vacuum decay. We have not seen this
phenomenon in our numerical simulations.}.

The outcome of the
entire process is thus the hole in the ghost condensate, a microscopic region
(of the size set by the length scale $M^{-1}$ and/or the UV cutoff) 
where the energy density is
negative, and the field strongly deviates from its background value. These
holes in the ghost condensate are likely to antigravitate.

\section{Nucleated bubbles}

The energy functional of the model is
\be
  E= \int~d^3x~[2P^{\prime}(X) \dot{\phi}^2 - P(X)]
\label{energy}
\ee
Besides the energy, there is another conserved quantity, the charge.
Indeed, the field equation has the form of current conservation,
\be
   \partial_\mu [ P^{\prime}(X)\partial^\mu \phi ] =0
\label{fieldeq1}
\ee
which implies the conservation of the charge
\be
   Q= \int~d^3x~P^{\prime}(X) \dot{\phi}
\label{charge}
\ee
It is precisely charge conservation that renders the background
(\ref{1*}) stable against small perturbations: even though for arbitrary
perturbation the energy can decrease,
small perturbations in the same charge
sector have larger energy
than the background itself.

\begin{figure}[htbp]
  \begin{center}
 \epsfig{file=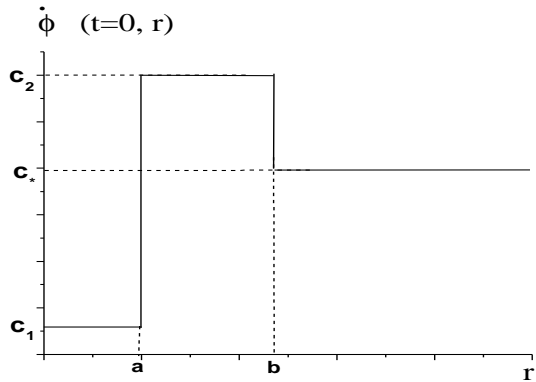,height=5cm,width=7cm} 
 \caption{An example of a configuration with vanishing energy and charge.}
  \label{fig-step}
  \end{center}
\end{figure}

\begin{figure}[htbp]
  \begin{center}
 \epsfig{file=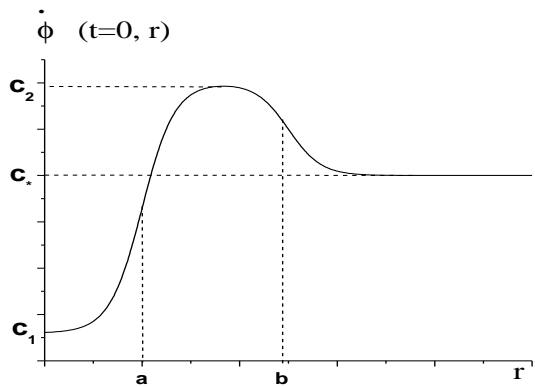,height=5cm,width=7cm}
  \caption{A bubble at the moment of nucleation.}
  \label{fig-smooth}
  \end{center}
\end{figure}

The latter property does not hold for large perturbations, so
the state (\ref{1*}) may tunnel into  other configurations
which have the same energy and charge. To be specific,
let us consider
the background
\be
       c= c_*
\label{c*}
\ee
(the generalization to $c > c_*$ is straightforward
provided that $c$ is not very large, otherwise one should
consider trial configurations with non-zero spatial gradient).
This configuration  has
\[
  Q=0 \; , \;\;\; E=0
\]
so final states the system can tunnel into have zero energy and charge.
An example of such a state is a 
spherically symmetric
configuration with zero spatial gradient,
$    \vec{\nabla} \phi = 0 $,
while the time derivative is a combination of step functions
(see fig.~\ref{fig-step})
\begin{eqnarray}
   \dot{\phi} &=& c_1 \;\; \mbox{at} \; 0<r<a \nonumber \\
   \dot{\phi} &=& c_2 \;\; \mbox{at} \; a<r<b \nonumber \\
   \dot{\phi} &=& c_* \;\; \mbox{at} \; r>b
\label{trial}
\end{eqnarray}
with $   0< c_1 < c_*$,  $ c_2 > c_* $.
The charge and energy of this configuration are
\begin{eqnarray}
\frac{3}{4\pi} Q &=& a^3 P^{\prime}(c_1^2)\cdot c_1
 +
(b^3 -a^3) P^{\prime}(c_2^2)\cdot c_2
\nonumber \\
\frac{3}{4\pi} E &=& a^3 [ 2 P^{\prime}(c_1^2) \cdot c_1^2 - P(c_1^2)] 
\nonumber \\
& & +
 (b^3-a^3) [ 2 P^{\prime}(c_2^2) \cdot c_2^2 - P(c_2^2)]
\nonumber
\end{eqnarray}
The first term in $Q$ is negative, while the second is positive
(this is why the region with $c_2 > c_*$ is needed). Requiring that
$Q=0$ one finds
\[
      b^3 - a^3 = -a^3\cdot
\frac{ P^{\prime}(c_1^2)\cdot c_1}{P^{\prime}(c_2^2)\cdot c_2}
\]
so one obtains for the energy
\[
E =  -a^3\cdot
P^{\prime}(c_1^2)\cdot c_1 \cdot [F(c_2) - F(c_1)]
\]
where
\[
  F(c) = \frac{ 2 P^{\prime}(c^2) \cdot c^2 -
P(c^2)}{P^{\prime}(c^2)\cdot c}
\]
For wide class of potentials, including
(\ref{P}), the function $F$ is positive at all $c$, has
a minimum  at some point $c_{**}$
on the left of the minimum of $P$, i.e.,
 $0 < c_{**} < c_{*}$, and
diverges as $c \to +0$ or $c \to +\infty$.

As a side remark,
the minimum
of $F$ occurs at the point where
\[
   2P^{\prime \prime}(c_{**}^2) c_{**}^2 
+ P^{\prime} (c_{**}^2) =0 
\]
It is straightforward to see that in the absence of gravity,
small perturbations about  the homogeneous background (\ref{1*})
grow exponentially for $c_{**} < c < c_*$ (tachyonic region)
and oscillate for $c< c_{**}$ and $c> c_*$ (in the former
region they have negative energy).

For $c_1<c_{**}$ (that is, necessarily
left of the tachyonic region) and sufficiently close to $+0$,
there exists
a point $c_2 > c_*$ such that
\[
  F(c_2) = F(c_1)
\]
With this choice of $c_1$ and $c_2$, both  energy and charge are
indeed equal to zero. The system can tunnel into such a configuration.
Further evolution will lead to non-zero spatial gradient of $\phi$,
because $\dot{\phi}$ is inhomogeneous in space for the trial
configuration just constructed. Note that only the ratio
$a/b$ is determined by the energy and charge, while
the
overall
spatial scale  remains a free parameter; this means that the
conservation of energy and charge allow for creation of bubbles of
arbitrary size.

It is worth noting that the step-function configuration of 
fig.~\ref{fig-step} has
been chosen
for illustration purposes only. The nucleated bubbles need not  have
steep walls. In what follows we have in mind smooth bubble profile at
the moment of nucleation, as shown in
fig.~\ref{fig-smooth}.

The initial and final states of tunneling are somewhat
unconventional here, since the field configurations
are time-dependent. Also, when describing tunneling, one 
has to impose a
constraint of charge conservation. The semiclassical formalism relevant
in this situation was in fact developed some time
ago~\cite{Lee:1988ge,Coleman:1989zu}. The bottom line is that
one perfoms the Wick rotation to the Euclidean space-time,
and considers pure imaginary fields there,
\be
     \phi = - i \theta
\label{im}
\ee
where $\theta$ is real. The boundary conditions at initial and final
Euclidean time, $\tau_i = -\infty$ and $\tau_f = 0$, respectively,
require that the field be homogeneous in space,
\be
         \vec{\nabla} \theta (\tau_{i,f}) = 0
\label{bc}
\ee
the latter condition being trivially satisfied
for the initial state (\ref{1*}),
since at large negative times $\theta (\vec{x}, \tau)
= i \phi (\vec{x}, t = -i\tau) = c\tau$. Generally speaking,
the semiclassical
tunneling exponent equals  the Euclidean action evaluated on
a solution to the Euclidean field equations obeying the
relations
(\ref{im}) and (\ref{bc}). In the model at hand, the
Euclidean action is still given by eq.~(\ref{1**}),
where now
\[
    X = (\partial_\mu \theta)^2
\]
and metric is Euclidean.

It is straightforward to see, however, that the model (\ref{1**}) as it
stands does not have relevant Euclidean solutions. Indeed,
under rescaling
\[
\theta (x^\mu) \to \lambda \theta (\lambda^{-1}x^\mu)
\]
the boundary conditions (\ref{bc}) and the
initial configuration $\theta = c \tau$  do not change, while
the action scales as
\[
   S \to \lambda^{4} S
\]
Thus, there are no relevant saddle points;
configurations of small size have small Euclidean action.

The scaling argument tells that the bubble nucleation rate is
UV dominated. If the UV cutoff $\Lambda$ is smaller than $M$, the decay
exponent is large,
\be
S \sim \left(\frac{M}{\Lambda} \right)^4
\label{est}
\ee
so the decay rate is
small\footnote{One way to understand this property 
is to consider
higher order operators added to
  the action (\ref{1**}). With a single additional
operator of the form
$\Lambda^{-(2n-6)} (\partial^n \phi)^2$  with $n>3$, the same
scaling argument gives the estimate
$S \sim (M/\Lambda)^{\frac{4n-12}{n-1}}$
for the action at a saddle point.
 Adding all higher order
operators effectively corresponds to the
limit $n \to \infty$, in which
the estimate (\ref{est}) is obtained.}, in accord with the analysis
in Ref.~\cite{AH}. 

\section{Real time evolution}

\subsection{Classical evolution in (1+1)-dimensions}

\subsubsection{Legendre transformation}
\label{ltransf} 

To understand how configurations of the type shown in
fig.~\ref{fig-smooth} 
evolve in
Minkowski space-time, let us first neglect the spatial
curvature of the bubble walls, and
thus
consider the system with the action (\ref{1**}) in $(1+1)$ dimensions.
The explicit form of
the field equation (\ref{fieldeq1}) in $(1+1)$ dimensions is
\be
(P^{\prime} + 2P^{\prime \prime}\phi_{,t}^2) \phi_{,tt}
- 4 P^{\prime \prime} \phi_{,x} \phi_{,t} \phi_{,xt}
-(P^{\prime}- 2 P^{\prime \prime} \phi_{,x}^2) \phi_{,xx} = 0
\label{ephi}
\ee
Here and in the following comma denotes differentiation.
This non-linear equation
is
simplified  by the
Legendre transformation. Instead of $x$, $t$, one chooses
new independent variables
\bea
  \xi &=& \phi_{,t} \nonumber \\
  \eta &=& \phi_{,x}
\label{deriv}
\eea
and, instead of $\phi$, one considers an unknown function
\[
  u (\xi, \eta) = - \phi + t\xi  + x \eta
\]
This change of variables is legitimate in any region where
the pairs
of coordinates $(\xi,\eta)$ and $(x,t)$
are in one-to-one correspondence.
 Equivalently, in such
a region
the Jacobian of the coordinate transformation,
\be
    J = u_{,\xi \xi} u_{,\eta \eta} - (u_{,\xi \eta})^2
 \label{Jacobian}
\ee
is  nowhere zero or infinite.
The second derivatives are
\bea
   \phi_{,tt} &=& J^{-1} \cdot u_{,\eta \eta} \nonumber \\
   \phi_{,xt} &=& - J^{-1} \cdot u_{,\xi \eta} \nonumber \\
    \phi_{,xx} &=& J^{-1} \cdot u_{,\xi \xi}
\eea
while the original coordinates are related to $\xi$ and $\eta$
as follows,
\bea
 t &=& u_{,\xi}
\label{coord1} \\
 x &=& u_{,\eta}
\label{coord}
\eea
The original equation (\ref{ephi}) in terms of new variables
takes the form
\be
     (P^{\prime} + 2P^{\prime \prime} \xi^2) u_{,\eta \eta}
      + 4 P^{\prime \prime} \xi \eta u_{,\xi \eta}
     -(P^{\prime}-2P^{\prime \prime} \eta^2) u_{,\xi \xi} = 0
\label{euxi}
\ee
where $P$ is a function of
the combination $(\xi^2 - \eta^2)$. This is now a linear equation
with coefficients that depend on the coordinates $\xi$, $\eta$.
Once a solution $u(\xi, \eta)$ is found, the quantities of interest
$\phi_{,t}\equiv \xi (x,t)$ and $\phi_{,x} \equiv \eta (x,t)$
are obtained, at least in principle,
by inverting eqs.~(\ref{coord1}), (\ref{coord}).

To simplify eq.~(\ref{euxi}), let us take advantage of its
Lorentz invariance and introduce new variables
$\rho$ and $\theta$ related to $\xi$ and $\eta$ as follows,
\bea
    \xi \equiv \phi_{,t} &=& \rho \cosh \theta
     \label{xi} \\
    \eta \equiv \phi_{,x} &=& \rho \sinh \theta
    \label{eta}
\eea
(of course, we made an assumption here that
      $ |\phi_{,t}| > |\phi_{,x}|$; it is valid in all cases
we consider).
Clearly, the meaning of $\rho$ is that it is a scalar
\[
   \rho = \sqrt{(\partial_\mu \phi)^2} \equiv \sqrt{X}
\]
In terms of the variables $\rho$ and $\theta$, the
original coordinates are
\bea
        t &=& u_{,\rho} \cosh \theta - \frac{1}{\rho}
       u_{,\theta} \sinh \theta
\label{trho} \\
       x &=& - u_{,\rho} \sinh \theta + \frac{1}{\rho}
       u_{,\theta} \cosh \theta
\label{xrho}
\eea
Now eq.~(\ref{euxi}) reads
\be
     \frac{P^{\prime}+ 2 P^{\prime \prime} \rho^2}{\rho^2}
u_{,\theta \theta}
-  P^{\prime} u_{,\rho \rho}
    - \frac{P^{\prime}+ 2 P^{\prime \prime} \rho^2}{\rho}
   u_{,\rho} = 0
\label{urho}
\ee
where $P=P(\rho^2)$, and prime still denotes the derivative with
respect to its argument $\rho^2$.

This equation is hyperbolic when $P^{\prime}$ and
$ (P^{\prime}+ 2 P^{\prime \prime} \rho^2)$
have the same sign, i.e., when the small perturbations of the field
$\phi$ do not grow (no tachyons). This equation is elliptic in
the tachyonic region, and has singularities (zeroes in front of
 one of the
second derivative terms) on the boundary of that region, i.e.,
at $P^{\prime}=0$ and at $ P^{\prime}+ 2 P^{\prime\prime} \rho^2 = 0$.
The second boundary
point is not of interest for our purposes; about the first boundary
point
we will have to say more later.

Let us consider the region on the
right of the minimum of $P$, where
both $P^{\prime}$ and $P^{\prime \prime}$ are positive.
The final simplification of eq.~(\ref{urho}) is made by introducing
a new variable $\tau$ instead of $\rho$,
\be
    \tau = \int_{c_*}^\rho~d\rho~
\sqrt{\frac{P^{\prime}+ 2 P^{\prime \prime} \rho^2}{P^{\prime}\rho^2}}
\label{taudef}
\ee
where for future convenience
the constant of integration is chosen in such a way that
\[
          \tau(\rho = c_*) = 0
\]
Then instead of eq.~(\ref{urho}) one obtains
\be
 u_{,\theta \theta} - u_{,\tau \tau}
-  F(\tau) u_{,\tau} = 0
\label{utau}
\ee
where
\[
     F(\tau) = \frac{\tau_{,\rho \rho}}{\tau_{,\rho}^2}
   + \rho \tau_{,\rho}
\]
It is instructive to point out that near the point $c_*$
one has
\[
     P^{\prime} = \mbox{const} \cdot (\rho^2 - c_*^2)
\]
Then the above expressions have simple form.
In particular,
\be
   \tau = 2\sqrt{\frac{\rho - c_*}{c_*}}
\label{tauapp}
\ee
and eq.~(\ref{utau}) reads
\be
   u_{,\theta \theta} - u_{,\tau \tau} - \frac{1}{\tau} u_{,\tau} = 0
\label{28a}
\ee
The latter equation has the form of
the wave equation in $(2+1)$ dimensions for
 $O(2)$-symmetric
functions.

\subsubsection{Shock waves}

\begin{figure}[htbp]
  \begin{center}
  \epsfig{file=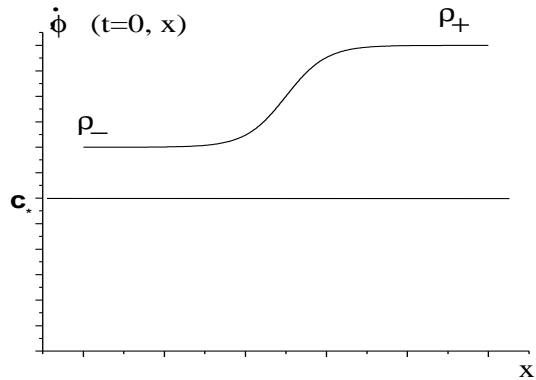,height=5cm,width=7cm}
 \caption{Initial data in the outer region of a bubble.}
  \label{fig-kink1}
  \end{center}
\end{figure}

Let us consider initial data such that $\phi_{,x} = 0$ everywhere
(cf. eq.~(\ref{bc})) and $\phi_{,t} > c_*$, i.e.,
$P^{\prime}$ and $P^{\prime \prime}$
are positive on
an entire line $-\infty <x < +\infty$.
Let us further assume that $\phi_{,t}$ at $t=0$ is a
monotonically increasing function of $x$, which tends to
certain finite values as $x \to \pm \infty$,
\bea
     \phi_{,t} (t=0, x\to -\infty) &\to& \rho_-
\label{phias-} \\
 \phi_{,t} (t=0, x\to +\infty) &\to& \rho_+
\label{phias+}
\eea
with
\[
     \rho_+ > \rho_- \geq c_*
\]
This initial configuration is shown in fig.~\ref{fig-kink1}, 
and it corrsponds
to the outer part of the bubble, namely, 
the region $r \gtrsim b$ for
the configuration of fig.~\ref{fig-smooth}. To this end,
$x$ is to be identified with $-r$
(thus, the bubble is meant to be on the right in fig.~\ref{fig-kink1};
this somewhat bizarre convention simplfies the further discussion).

These initial data are translated into the Legendre variables
as follows. According to eqs.~(\ref{eta}), (\ref{trho})  
and (\ref{xrho}),
the initial condition $\phi_{,x} =0$  is satisfied when the line
$t=0$ coincides with the line $\theta =0$, and $u=\mbox{const}$
on that line. Thus, the initial data are
\bea
         u_{,\theta} &\neq& 0 \; , \;\;\;\; \theta =0
\label{utheta0} \\
         u_{,\rho} &=& 0 \; , \;\;\;\; \theta=0
\label{urho0}
\eea
Once the initial data $\phi_{,t} (t=0,x)$ is specified,
the initial data for the Legendre variables is obtained by
inverting the relations
\bea
      \rho &=&  \phi_{,t} (t=0,x) \nonumber \\
        x &=& \frac{1}{\rho} u_{,\theta} (\theta =0, \rho)
\label{33}
\eea
According to eqs.~(\ref{xi}) and (\ref{phias-}), (\ref{phias+}),
the initial data for $u_{,\theta}$ are specified
on a finite interval
\be
    \rho_- < \rho < \rho_+
\label{intrho}
\ee
Since $\phi_{,t}$ at $t=0$ is an increasing function of $x$,
eq.~(\ref{33}) means that $u_{,\theta}$ at
$\theta = 0$ is an increasing function of $\rho$.
Since $x$  runs from $-\infty$ to $+\infty$,
eq.~(\ref{33}) implies that $u_{,\theta}$ is singular
at the ends of the interval (\ref{intrho}),
\bea
        u_{,\theta} (\theta = 0, \rho \to \rho_-)
       &\to& - \infty
\label{u-} \\
 u_{,\theta} (\theta = 0, \rho \to \rho_+)
       &\to& + \infty
\label{u+}
\eea
Instead of $\rho$, one can of course use the coordinate $\tau$
in the above expressions.

\begin{figure}[htbp]
  \begin{center}
  \epsfig{file=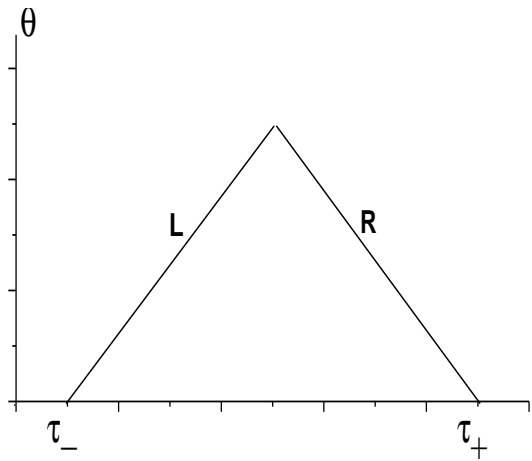,height=6cm,width=7cm}
 \caption{Singularity lines in $(\tau, \theta)$-plane.}
  \label{fig-triangle}
  \end{center}
\end{figure}

Let us now discuss the behavior of the solution
$u(\tau, \theta)$  to eq.~(\ref{utau}),
in the $(\tau, \theta)$ plane.
The singularities (\ref{u-}) and (\ref{u+})
that start at $\tau_-$ and $\tau_+$, respectively, move along the 
light-like
lines on this plane. The relevant light-like lines are the lines 
L and R
in fig.~\ref{fig-triangle} 
(the other two lines are screened by the  lines L and R).

The right singularity, labeled R in  fig.~\ref{fig-triangle}, occurs at
\be
        \theta = \tau_+ - \tau
\label{rsing}
\ee
Near the right singularity
the function $u$ has the form
\be
     u = A(\tau) C_+(\theta + \tau)
\label{uright}
\ee
where
$A(\tau)$ is a slowly varying function,
while $C_+(\bar{z} \equiv (\theta + \tau))$ varies rapidly and has the property
\[
       C_+ (\bar{z}) \to +\infty \; , \;\;\;
        \bar{z}  \to \tau_+ - 0
\]
The form of $C_+$ is determined by the singular part of
the initial configuration $u_{, \theta} (\theta =0, \rho)$
as $\rho \to \rho_+$, i.e., by the behavior of 
$\phi_{,t} (t=0, x)$ as $x \to + \infty$.

>From eqs.~(\ref{trho}) and (\ref{xrho}) one finds that
near the right singularity
\bea
       t &=& \frac{A(\tau)}{\rho(\tau)} C_+^{\prime}(\theta + \tau)
           (\rho\,\tau_{,\rho} \cosh \theta
- \sinh \theta)
\label{tright} \\
x &=& \frac{A(\tau)}{\rho(\tau)} C_+^{\prime}(\theta + \tau)
           (- \rho\,\tau_{,\rho} \sinh \theta +
\cosh \theta)
\label{xright}
\eea
Note that from eq.~(\ref{taudef}) it follows that
$ \rho\,\tau_{,\rho} > 1$,
so time $t$ is positive near the right singularity. The coordinate $x$
is also positive near the right singularity, at least in the lower right
corner of  fig.~\ref{fig-triangle}, 
where $\theta$ is small.
Thus, the lower right part of the line R corresponds to
the asymptotics $t \to \infty$, $x \to +\infty$.

\begin{figure}[htbp]
  \begin{center}
  \epsfig{file=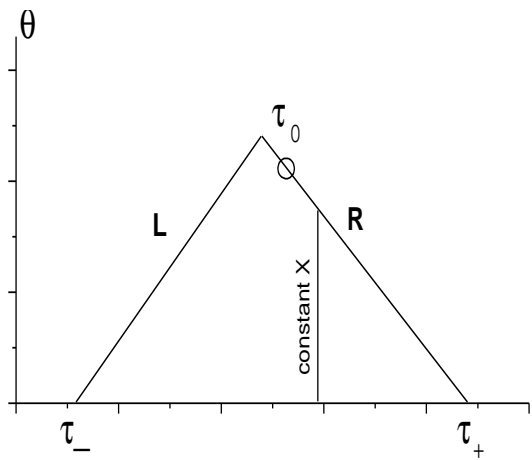,height=6cm,width=7cm}
 \caption{Line of constant $(\partial_\mu \phi)^2$
in $(\tau, \theta)$-plane.}
  \label{fig-constX}
  \end{center}
\end{figure}

A line of constant $\tau$, i.e., a line of constant
$X\equiv (\partial_\mu \phi)^2 $, shown in fig.~\ref{fig-constX},
hits the singularity. According to eqs.~(\ref{tright}) and (\ref{xright}),
 at large $t$ and $x$ this line
becomes
a straight line in $(x,t)$-plane, with the slope
\be
        \frac{x}{t} = \frac{\cosh (\tau_+ - \tau)
- \rho\,\tau_{,\rho} \sinh (\tau_+ - \tau)}{\rho\,\tau_{,\rho} \cosh
(\tau_+ - \tau) - \sinh (\tau_+ - \tau)}
\label{slope}
\ee
where we made use of the relation (\ref{rsing}) to express $\theta$
through $\tau$.
For $\tau$ sufficiently close to $\tau_+$, the lines of constant
$X$
move towards positive $x$, so there is a wave
moving right. In the case of the bubble of fig.~\ref{fig-smooth} 
this wave moves
{\it inwards} the bubble.

\begin{figure}[htbp]
  \begin{center}
  \epsfig{file=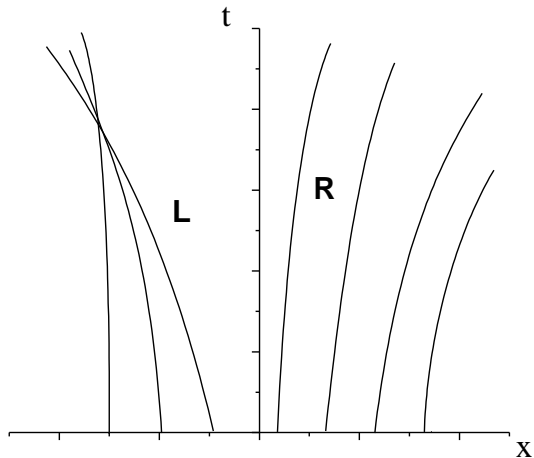,height=6cm,width=7cm}
  \caption{Lines of constant $X \equiv (\partial_\mu \phi)^2$ 
 in $(x,t)$-plane. Regions R and L correspond at large
$t$ to regions near the lines R and L in fig.~\ref{fig-triangle}.}
  \label{fig-xt}
  \end{center}
\end{figure}

For generic initial data
there is a point on the right singularity line R where
the two terms in the numerator
in eq.~(\ref{slope})
cancel each other (shown by a circle 
in fig.~\ref{fig-constX}). Then left of
this point (i.e., at $\tau < \tau_0$) the coordinate $x$ is negative
(and tends to $- \infty$ as the line of constant $\tau$ approaches
the singularity line). In this region,
the lines of constant
$(\partial_\mu \phi)^2 $ move left, so there is a left-moving wave.
In between the left-moving and right-moving waves,
the space is gradually occupied by ``constant'' ghost condensate
\bea
       \phi_{,t} &=& \rho(\tau_0) \cosh (\tau_+ - \tau_0)
\nonumber \\
        \phi_{,x} &=& \rho(\tau_0) \sinh (\tau_+ - \tau_0)
\eea
The final point to mention about the right singularity line is
that the slope (\ref{slope}) increases as $\rho$ increases,
so the lines of constant $X$ do not intersect
on $(x,t)$-plane, see fig.~\ref{fig-xt}. Indeed, one finds
\begin{eqnarray}
   \left(\frac{x}{t}\right)_{,\rho} &=&
\frac{2 \rho^3}{\sqrt{1
+ 2 P^{\prime \prime} \rho^2/P^{\prime}}}  
\nonumber \\
& & \times
\frac{1}{[\rho\,\tau_{,\rho} \cosh (\tau_+ - \tau)
- \sinh (\tau_+ - \tau)]^2} 
\nonumber \\
& & \times
\left (3 \frac{P^{\prime \prime 2}}{P^{\prime 2}} -
\frac{P^{\prime \prime\prime}}{P^{\prime}}
\right)
\label{sl}
\end{eqnarray}
which is positive for large class of the potentials $P$.
As it evolves, the wave gets spread over larger interval of $x$.
The  asymptotic
velocity of the wave is always smaller than the speed
of light: the maximum slope (\ref{slope}) is attained in the lower
right corner where $\tau \approx \tau_+$ and $\theta \approx 0$, and
there
\[
  \frac{x}{t} = \frac{1}{(\rho\,\tau_{,\rho})(\tau_+)} < 1
\]
For initial data close to the minimum of $P$,
the maximum slope, and hence the velocity of the wave are small,
since $\rho\,\tau_{,\rho}$ is large, see eq.~(\ref{tauapp}).

Let us turn to the left singularity line L.
This line corresponds to
\[
    \theta = \tau - \tau_-
\]
and the solution near the singularity line has the form,
cf. eq.~(\ref{uright}),
\be
     u = A(\tau) C_- (\tau - \theta)
\ee
The function $C_-$ is rapidly varying;
it is an increasing function of its argument
with the property
\[
        C_- (z) \to -\infty \; , \;\;\;\;
        z \equiv (\tau - \theta) \to \tau_- + 0
\]
At the very first sight the
situation here is similar to that near the line R. This is not
quite true,
however. The expressions for $t$ and $x$ near the line L
(analogs of eqs.~(\ref{tright}) and (\ref{xright})) are
\bea
       t &=& \frac{A(\tau)}{\rho(\tau)} C_-^{\prime}( \tau - \theta)
           (\rho\,\tau_{,\rho} \cosh \theta
+ \sinh \theta)
\label{tleft} \\
x &=& - \frac{A(\tau)}{\rho(\tau)} C_-^{\prime}(\tau - \theta)
           ( \rho\,\tau_{,\rho} \sinh \theta + \cosh \theta)
\label{xleft}
\eea
Time $t$ is again positive
while $x$ is now  negative.
Thus the left part of fig.~\ref{fig-triangle} describes a wave
moving left. In the context of a bubble of fig.~\ref{fig-smooth}
this wave moves outwards.

The point, however, is that the absolute values of
the slopes of lines of constant $\tau$ (i.e., constant $X$)
on the $(x,t)$-plane
{\it increase} as $\rho$ increases: the relevant formula
is again given by eq.~(\ref{sl}), but with the
opposite overall sign (the wave moves left)
and with $(\tau - \tau_-)$ substituted
for $(\tau_+ - \tau)$.
Thus, at given large $t$,
points (in $x$-space) with larger $\rho$ are to the left of  points
with smaller $\rho$. On the other hand, at $t=0$ the situation is 
opposite, since the initial data are such that $\phi_{,t} = \rho$
increases with $x$ as shown in fig.~\ref{fig-kink1}. This means that lines of
constant $X$ intersect on $(x,t)$-plane, see fig.~\ref{fig-xt}.
Of course, the whole treatment breaks down when these lines intersect,
as there emerges a singularity
on $(x,t)$-plane. This singularity is of the type of
shock wave,
or kink,
at which the first derivatives $\phi_{,t}$ and $\phi_{,x}$
are step functions: just before the lines of
constant $X$ intersect, the values of $X(x)$, and hence 
$\phi_{,t}$ and $\phi_{,x}$ are substantially different at
neighboring points of space. From the above analysis
it follows that this shock wave moves left in fig.~\ref{fig-kink1}, 
at least at the moment
when it gets formed --- in the
context of a bubble this is the motion outwards.

\subsubsection{Numerical analysis}

Equation~(\ref{ephi}) for the evolution of the ghost field can be
studied numerically.  For the purpose, we found it convenient
to take $\xi=\phi_{,t}$ and $\eta= \phi_{,x}$ as new dependent field
variables\footnote{Here we consider $\xi$ and $\eta$ still as
dependent functions of $t$ and $x$, i.e.~we do not perform the
Legendre transformation of subsection \ref{ltransf}.}.  In
terms of these, the equation of motion can be recast into the
form
\be
(P^{\prime} + 2P^{\prime \prime}\xi^2) \xi_{,t}
- 4 P^{\prime \prime} \eta \xi \xi_{,x}
-(P^{\prime}- 2 P^{\prime \prime} \eta^2) \eta_{,x} = 0
\label{num1}
\ee
where $P=P(\xi^2-\eta^2)$,
with the additional constraint
\be
\eta_{,t}=\xi_{,x}
\label{num2}
\ee
We discretize now the space dependence of the variables $\xi, \eta$
by introducing a lattice with uniform lattice spacing $a$ and defining
\bea
\xi_n &=& \xi (na) \nonumber \\
\eta_n &=& \eta (na) \label{num3}
\eea
We also make the overall lattice finite imposing $-N \le n \le N$,
introduce Neumann boundary conditions on $\phi$, and replace
the derivatives $\xi_{,x}, \eta_{,x}$ with their central difference
approximations $(\nabla \xi)_n=(\xi_{n+1}-\xi_{n-1})/(2a),
(\nabla \eta)_n=(\eta_{n+1}-\eta_{n-1})/(2a)$.
After these steps eqs.~(\ref{num1}), (\ref{num2})
take the form of a set of coupled ordinary differential equations
of the first order for the time dependence of the variables
$\xi_n, \eta_n$:
\bea
\frac{d\xi}{dt} &=&
\frac{4 P^{\prime \prime} \eta \xi \nabla \xi
+(P^{\prime} - 2 P^{\prime \prime} \eta^2) \nabla \eta}
{(P^{\prime} + 2P^{\prime \prime}\xi^2)}
\label{num4} \\
\frac{d\eta}{dt} &=& \nabla \xi
\label{num5}
\eea
We integrated these equations in time by using the
second order Runge-Kutta formula.

We would like to make a few observations about our numerical procedure.
The main concern in integrating evolution equations such as the ones
at hand is the possible onset of numerical instabilities.  The avoidance
of such instabilities limits the maximum time step to values that
are generally small enough to render the use of higher order
integration formulae unwarranted.  This is why we used the second
order Runge-Kutta algorithm.  The adequacy of this algorithm was
confirmed by the conservation of energy, which, in the stability
region, was accurate to better that one part per million.
However, the time integration can become unstable.  This happens
a) when the formation of a shock wave generates a wave-front
spanning only a few lattice spacings and b) when the signature
of the equation is elliptic over some domain of values of $x$.
In the case (a) the approximation to the space derivative obviously
fails, and this leads to instabilities in the time integration.
We have noticed, incidentally, that the onset of such instabilities
is more serious if one uses $\phi$ itself, rather than $\eta=\phi_{,x}$,
as one of the independent variables.  The use of $\eta$ avoids
the need of discretizing second derivatives with respect to
$x$ and leads to a better behaved integration algorithm.
 From the physical point of view, one can also argue that
$\phi_{,x}$ plays a more fundamental role than $\phi$ in the
system under consideration.

The instabilities that occur in the case (b), namely in the domains
where the equation is elliptic, are of physical 
nature, and not caused by shortcomings of the integration procedure.
Indeed, with an elliptic equation, perturbations grow
exponentially when one integrates the equation forward in time.
The damping of such instabilities would be left to the UV
completion of the theory.

Whether due to a shock wave or to a domain of elliptic
signature, numerical instabilities are characterized by an
exponential growth of the short range fluctuations.
In order to tame such instabilities we supplemented our evolution
algorithm with an additional step, which serves to dampen the
growth of those fluctuations.  Namely, after the standard
Runge-Kutta integration step, we use a fast Fourier transform (FFT)
to calculate
\bea
\tilde \xi_k &=& \sum_n e^{2  \pi \imath k n/N}\, \xi_n
\nonumber \\
\tilde \eta_k &=& \sum_n e^{2 \pi \imath k n/N}\, \eta_n
\label{num6}
\eea
and dampen $\tilde \xi, \tilde \eta$ by replacing
\bea
\tilde \xi_k &\to& \tilde \xi_k' =
e^{-\alpha [1-\cos(2 \pi k/N)]\, dt/a^2}\,\tilde \xi_k
\nonumber \\
\tilde \eta_k &\to& \tilde \eta_k' =
e^{-\alpha [1-\cos(2 \pi k/N)]\, dt/a^2}\,\tilde \eta_k
\label{num7}
\eea
$dt$ being the integration time step and $\alpha$ suitable parameter.
After this step, we go back to the original variables by the inverse
FFT.

We have found that,
with a judicious choice of the damping factor $\alpha$,
we were able to evolve the equations well past the formation of
the shock wave and also for some time
in the presence of an elliptic domain,
while preserving the conservation of energy to a few per cent,
or better.  Of course, the front of the shock gets slightly smoothed
out and, in the elliptic domain, the short range fluctuations,
after some initial wild growth, 
get gradually damped down and become intolerable only at relatively late
times.
It is interesting to observe that our procedure, while mainly
motivated by the desire to keep the numerical integration under
control, could also be thought of as a way to mimic the effects
of the UV completion.

Finally, the discretization of the equations for the three-dimensional
evolution with spherical symmetry proceeds along the same lines,
after having inserted the appropriate metric factors (powers of $r$).

\begin{figure}[htbp]
  \begin{center}
  \epsfig{file=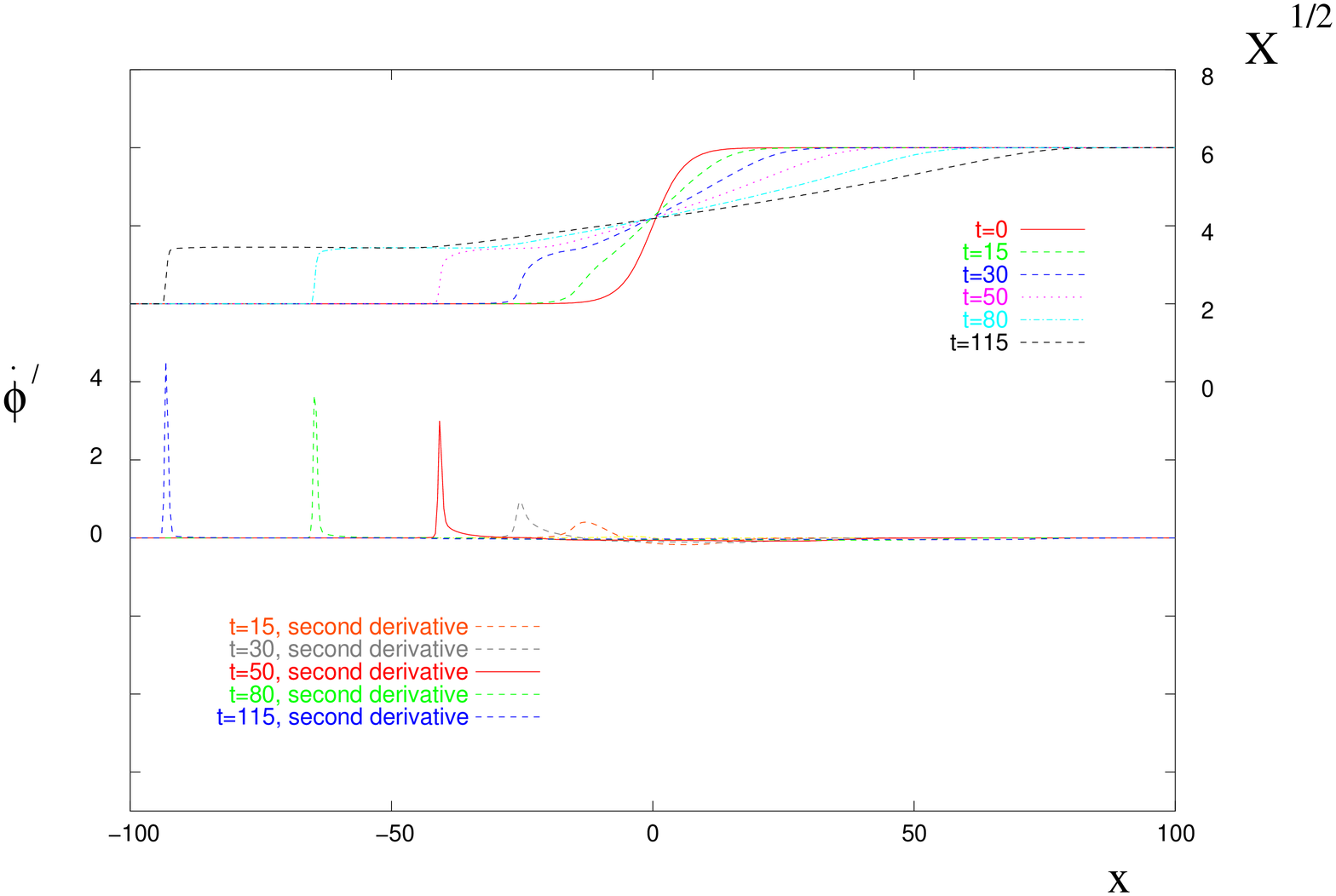,height=13cm,width=14cm}
  \caption{(1+1)-dimensional system.
Profiles of $\sqrt{X} \equiv \sqrt{\dot{\phi}^2 - \phi^{\prime 2}}$
(upper part) and $\dot{\phi}^{\prime}$ are shown at different times.
Scales of $x$ and $t$ are arbitrary.}
  \label{kink1d}
  \end{center}
\end{figure}

In fig.~\ref{kink1d} we show
the profile of the field configuration
obtained by solving eq.~(\ref{ephi}) numerically, with 
the potential $P$ given by eq.~(\ref{P}) and initial data
similar to those shown in fig.~\ref{fig-kink1}. This solution
clearly demonstrates the formation of the shock wave and its motion to
the left. There is also a soft wave slowly propagating to the
right. In our numerical simulations we always observed this behavior
of solutions for initial data of the type shown in fig.~\ref{fig-kink1}.

\subsubsection{Around the critical point}

\begin{figure}[htbp]
  \begin{center}
  \epsfig{file=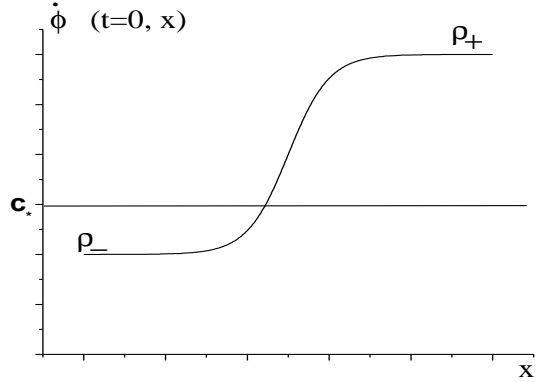,height=5cm,width=7cm}
 \caption{Initial data in the interior of a bubble.}
  \label{fig-kink2}
  \end{center}
\end{figure}

Let us now consider another initial configuration,
again with $\phi_{,x} = 0$ but with the profile of $\phi_{,t}$
shown in fig.~\ref{fig-kink2}. This configuration corresponds to
the region $r \loe a$ in fig.~\ref{fig-smooth}, i.e.,
the region near the point where $ \phi_{,t} = c_*$
(in fig.~\ref{fig-kink2} the bubble is meant to be on the left,
like in fig.~\ref{fig-smooth}, and unlike in fig.~\ref{fig-kink1}).
At the left of the critical point $ \phi_{,t} = c_*$ the system is unstable.
Let us study whether this instability propagates into the
region on the right of this critical point, and also study
the motion of this critical point as time evolves.

Let us consider the behavior of the 
solution $\phi(x,t)$ in the region where $X \geq c_*^2$. In
this region one 
can still use the variables $\rho$ and $\theta$ (or, equivalently, $\tau$ and
$\theta$) in the Legendre conjugate problem. Now, the critical point
$X\equiv\rho^2=c_*^2$ corresponds to $\tau=0$, and in the vicinity of this
point the function $u(\tau,\theta)$ obeys eq.~(\ref{28a}). 
>From the analogy to
$O(2)$-symmetric $(2+1)$-dimensional problem it is clear that the solution and
its derivatives never become singular at $\tau=0$, provided initial data are
smooth. Futhermore, the solution is uniquely determined by the initial data
at $X\geq c^2_*$. This means that the behavior of the solution
$\phi(x,t)$ in the region $X\geq c^2_*$ is not sensitive to its behavour
at $X<c^2_*$; the regions right of the critical point and left of the
critical point do not talk to each other; whatever happens in the inner,
tachyonic region of the bubble, has no effect on the outer 
region\footnote{In the classical
field theory problem, a singularity may eventually be formed
in the tachyonic region, and the solution does not globally exist
at later times.}.

\begin{figure}[htbp]
  \begin{center}
 \epsfig{file=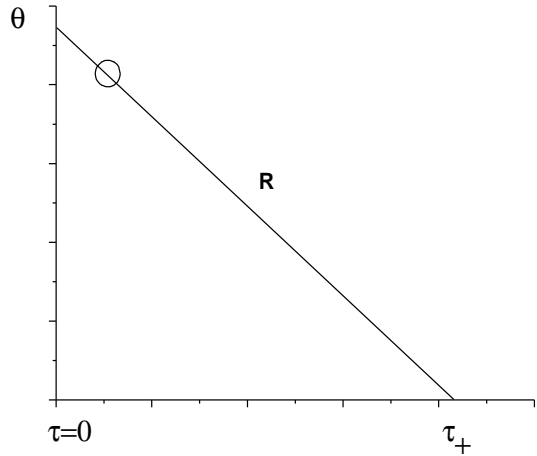,height=6cm,width=7cm} 
 \caption{Singularity line in $(\tau, \theta)$-plane for initial
data of fig.~\ref{fig-kink2}.}
  \label{fig-triangle2}
  \end{center}
\end{figure}

  Let us see that the critical point $X=c^2_*$ moves left in 
fig.~\ref{fig-kink2}, i.e.,
inwards the bubble. To this end, we note that the initial data for the
Legendre problem are again formulated in a finite interval of $\rho$, which
is now $c_* \leq \rho <\rho_+$, i.e., in a  finite interval 
$0 < \tau< \tau_+$. The initial data is $u(\theta=0,\tau)=0$, while
$u_{,\theta} (\theta=0,\tau)$ is singular at $\tau=\tau_+$. At non-zero
$\theta$, the singularity propagates along the light-like line
$\theta=\tau_+ - \tau$, as shown in fig.~\ref{fig-triangle2}. 
Near the singularity line the
solution still has the form given by eq. (\ref{uright}), 
and this line again
corresponds to $t\rightarrow\infty$. The slope (\ref{slope}) 
is positive on the lower
right part of this line where $(\tau_+ -\tau)$ is small, so the wave
corresponding to the region near the corner $\theta=0$, $\tau=\tau_+$ is
again moving right in fig.~\ref{fig-kink2}. 
This is now the motion outwards the bubble.

At $\rho$ close to $c_*$ one has $\rho\,\tau_{,\rho}=(\rho/c_*-1)^{-1/2}$,
see eq.(\ref{tauapp}), 
which is large. Thus there is always a point denoted by a circle 
on the
singularity line $R$, at which the slope (\ref{slope}) 
vanishes. Regions right and
left of this point correspond to motion right and left in 
fig.~\ref{fig-kink2}, i.e.
outwards and inwards the bubble, respectively. 
In particular, the critical 
point $\tau=0$ (i.e., $X=c_*^2$)
moves left in fig.~\ref{fig-kink2}, i.e., inwards the bubble.

Another, 
direct way to see that the regions with $X<c_*^2$ and $X>c_*^2$
do not communicate with
each other is to consider eq.~(\ref{ephi}) itself.
Rescaling the field in such a way that 
\[
     c_* = 1
\]
one has near the critical point 
\[
 P^{\prime} = X-1
\]
Equation (\ref{ephi}) then reads
\be
  (3 \phi_{,t}^2 - \phi_{,x}^2 -1) \phi_{,tt}
 - 4 \phi_{,x} \phi_{,t} \phi_{,xt}
 - ( \phi_{,t}^2 - 3\phi_{,x}^2 -1) \phi_{,xx} = 0
\label{cl1}
\ee
This may be viewed as a wave equation with coefficients depending on
$\phi_{,x}$ and $\phi_{,t}$. Let us find the characteristics of this
equation that starts at a point $(x,t)$ where 
$X \equiv \phi_{,t}^2 - \phi_{,x}^2 = 1$, i.e., at the critical point.
At this point, eq.~(\ref{cl1}) reduces to
\be
  \phi_{,t}^2  \phi_{,tt}
 - 2 \phi_{,x} \phi_{,t} \phi_{,xt}
 + \phi_{,x}^2  \phi_{,xx} = 0
\label{cl2}
\ee
Thus, the equation for the characteristics is
\[
  \phi_{,x}^2 \cdot dt^2 + 2 \phi_{,x} \phi_{,t} \cdot dx dt
  + \phi_{,t}^2 \cdot dx^2 =0
\]
We see that the characteristics are degenerate and obey
\be
 \frac{dx}{dt} = - \frac{\phi_{,x}}{\phi_{,t}}
\label{cl3}
\ee
Let us now see that the critical point where
$X=1$ moves precisely along
this characteristic.
The motion of the critical point is determined by the equation
\[
 d(\phi_{,t}^2 - \phi_{,x}^2 -1 ) \equiv
(\phi_{,t}^2 - \phi_{,x}^2 -1 )_{,t} dt +
 (\phi_{,t}^2 - \phi_{,x}^2 -1 )_{,x} dx = 0
\]
which gives
\[
\frac{dx}{dt} = 
- \frac{\phi_{,t} \phi_{,tt} - \phi_{,x} \phi_{,xt}}{\phi_{,t} 
\phi_{,xt} - \phi_{,x} \phi_{,xx}}
\]
Making use of eq.~(\ref{cl2}) we find that the right hand side of 
this expression is precisely the same as  
the right hand side of eq.~(\ref{cl3}),
so the charactristic and the world line of the point where
$X=1$, indeed
coincide.

Since the line $X=1$ is a characteristic, signals emitted on the left
of
this line do not propagate to the right of this line, and vice versa,
i.e., the regions with $X<c_*^2$ and $X>c_*^2$ indeed
evolve independently. Equation (\ref{cl3}) determines the motion of
the critical point; for an initial configuration shown in 
fig.~\ref{fig-kink2} the
velocity of this point initially vanishes (since $\phi_{,x} = 0$
at $t=0$), and then becomes negative, as positive $\phi_{,x}$
develops.
The critical point indeed moves inwards the bubble.

One way to understand the unusual property that the evolution
in the regions with $X< c_*^2$ and $X>c_*^2$ proceeds independently,
is to recall the expressions for the energy-momentum tensor and conserved
current,
\begin{eqnarray}
  T_{\mu \nu} &=& 2 P^{\prime}(X) \partial_\mu \phi \partial_\nu \phi
- P(X) \eta_{\mu \nu}
\label{tei} \\
 j_\mu &=& P^{\prime}(X) \partial_\mu \phi
\end{eqnarray}
One observes that all components of $T_{\mu \nu}$ and
$j_\mu$ are equal to zero at the surface in space where $X=c_*^2$.
Thus, there is no transfer of energy and charge between the regions
with $X<c_*^2$ and $X> c_*^2$. This explains, at least partially,
the independence of the evolution in these two regions. It is worth
stressing that this
argument works in any number of dimensions.

\begin{figure}[htbp]
  \begin{center}
 \epsfig{file=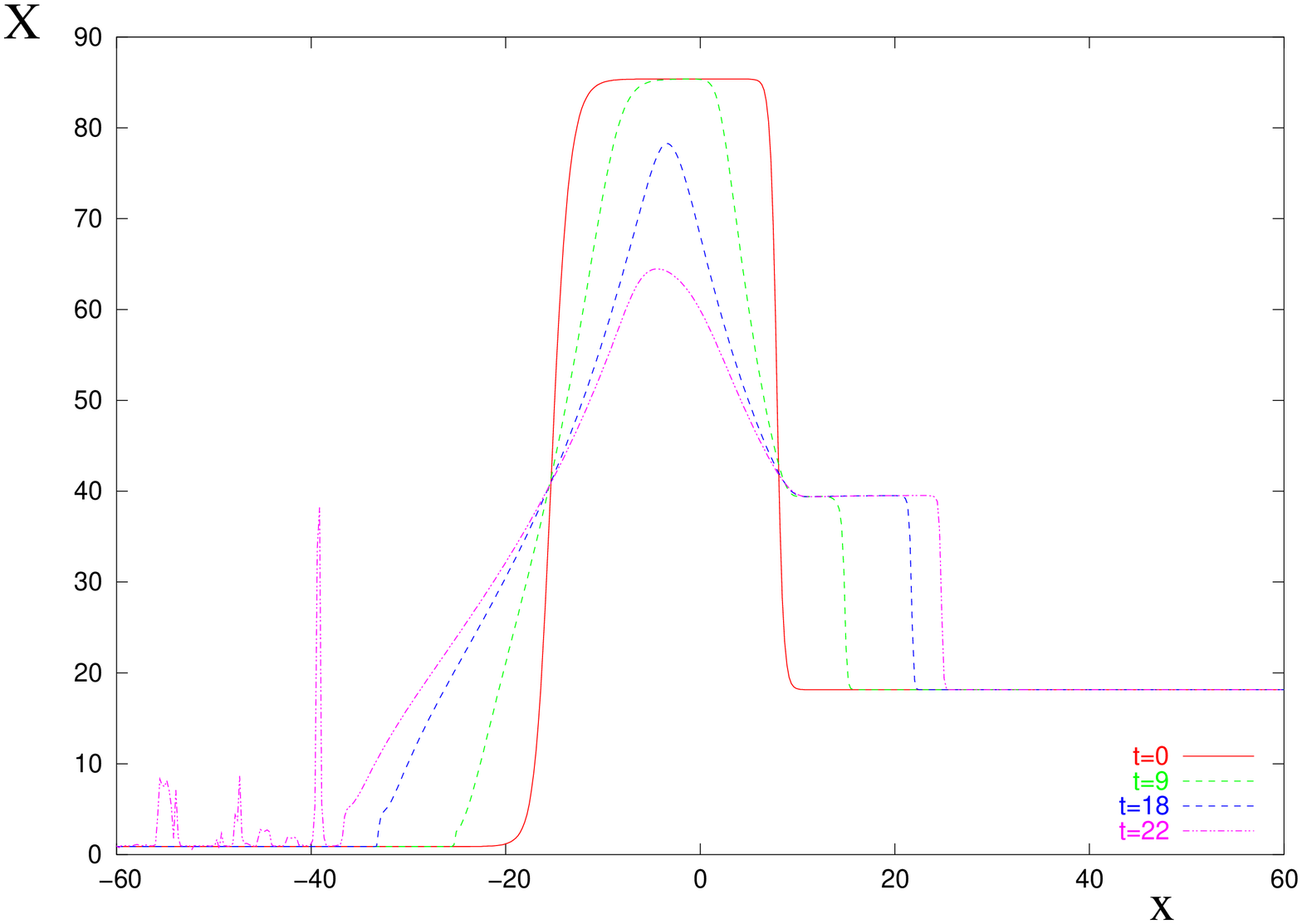,height=8cm,width=12cm} 
 \caption{Initial stages of evolution of a bubble in $(1+1)$
dimensions. Profiles of $X$ are shown
at different times. The irregular pattern
at  $x \loe -40$ for $t=22$ is due to instabilities in the 
tachyonic region where $X< c_*^2 \equiv 1$.}
  \label{evol-1d}
  \end{center}
\end{figure}

\subsubsection{Overall evolution} 

Combining the two pictures, corresponding to the initial data of 
figs.~\ref{fig-kink1}
and \ref{fig-kink2}, 
we obtain the following qualitative properties of the evolution of
a bubble, still in $(1+1)$ dimensions. The inner part of the bubble
is unstable due to the
tachyonic
character of the region $c_{**}^2<X<c_*^2$. Nevertheless, the region 
$X\geq c_*^2$ is not sensitive to this instability.
The critical point where $X=c_*^2$ moves
towards the center of the bubble, so the region of large fields 
(hole in the ghost
condensate) remains of small size.
In the outer part, a wave moving outwards the bubble develops. Eventually
this wave forms a kink. We show in 
fig.~\ref{evol-1d}  the
evolution of a bubble, which we obtained by solving 
eq.~(\ref{ephi}) numerically
with the bubble-like initial data (in particular, initially 
$\phi_{,x}=0$). The region left of $x \approx - 20$ in this figure
(units of $x$ and $t$
are arbitrary) has initially $X < 1$. The field at large
negative $x$ is thus unstable; this instability is manifest in 
fig.~\ref{evol-1d}. The motion of the boundary of the tachyonic
 region to the
left
(inwards the bubble) and a shock wave moving right are also clearly 
visible.

\subsection{Four dimensions}

Although the above analysis has been performed in
$(1+1)$ dimensions, we argue that our main findings 
are valid in $(3+1)$-dimensional theory as well, at least for 
$O(3)$-symmetric bubbles. At large distances from the bubble center,
the curvature of a sphere $r=\mbox{const}$ is small, so the 
evolution is similar to that in $(1+1)$-dimensional theory.
There are two consequences of this simple observation. First,
the tachyonic region, where $X<c_*^2$, does not expand; instead, it
shrinks and forms a hole in the ghost condensate. The energy of this
region is negative but finite at the moment of nucleation. 
According to eq.~(\ref{tei}), there is no transfer of energy from the
hole to the outer region where $X>c_*^2$, so the overall energy of the
outer region, referenced from the energy of the original background
$\dot{\phi}=c$, remains a finite constant\footnote{Assuming that the
UV effects render the energy of a microscopic hole finite, the latter
property is independent of the form (\ref{tei}) of the energy-momentum
tensor, which is not exact in a complete theory.}.  
  
Second, an outgoing wave of positive energy eventually forms a shock.
It is conceivable that the UV effects smoothen out the profile of the
wave. In front of the wave, i.e., at large $r$, the ghost field is
still in its original state, $\vec{\nabla}\phi =0$, $\dot{\phi} = c$.
Let us discuss the field behind the wave.

Let us first consider the special value of
the original background, $\dot{\phi} =
c_*$. In that case the conservation of energy requires that the
background behind the outgoing wave is also $\dot{\phi} = c_*$,
$\vec{\nabla} \phi=0$. This follows from the fact 
that the energy density
of the original background is zero, while the overall energy of the
positive energy region (outside the hole) is finite. This implies that
$X=c_*^2$ behind the outgoing wave, otherwise the energy of the
positive energy region would increase as $r^3(t)$ where $r(t)$ is the
radius of the wave, which grows in time, $r(t) \to \infty$ 
as $t\to\infty$. The property that $X=c_*^2$ behind the wave, 
does not guarantee by itself 
that the background behind the wave is the same as
the original background. However, the energy density of the wave
decreases in time (otherwise its total energy would grow as
$r^2(t)$). Therefore the amplitude of the wave decays, which is
only possible if $\vec{\nabla} \phi =0$, $\dot{\phi}= c_*$ both in
front of and behind the wave\footnote{This 
argument does not work in $(1+1)$ dimensions:
the conservation of energy and charge do not forbid that
the background behind the outgoing wave is Lorentz-boosted 
with respect to the initial background. We have not observed such a
phenomenon in our numerical simulations.}.    

The above argument does not work for the general values of
the original background, 
$\dot{\phi} = c > c_*$. In that case the conservation of energy and
charge does not forbid that the energy density behind the
outgoing wave  be smaller than the energy density of the original
 background (i.e., of the field in front of the wave).
However, in our numerical simulations the background behind the outgoing
wave(s) always settled down to its original value, so the energy of the
wave did  not grow in time, and the amplitude of the wave 
decreased as it moved away from the bubble center.

\begin{figure}[ht!]
\begin{center}
\epsfig{file=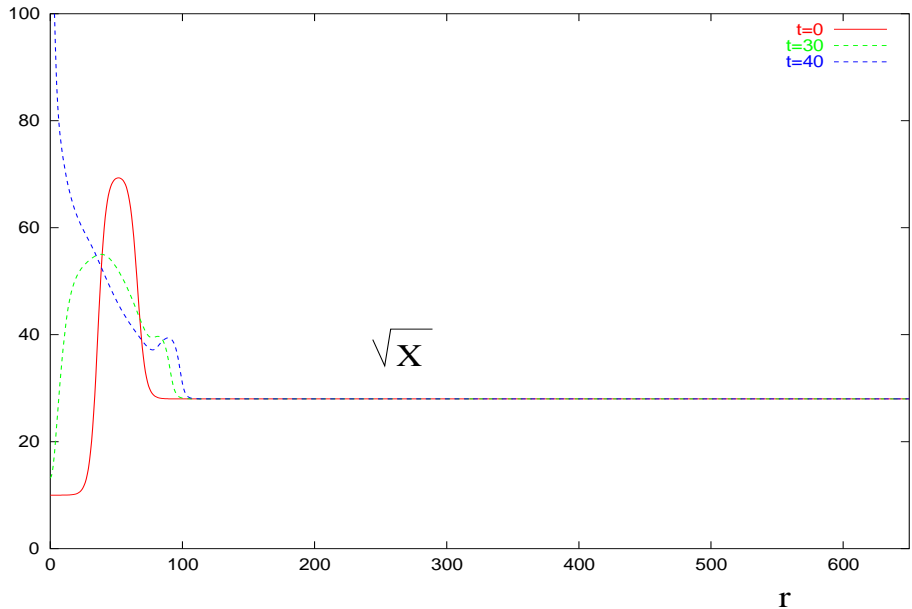,height=8cm,width=12cm} 
\caption{Initial stages of evolution of a bubble in $(3+1)$
dimensions. Profiles of $\sqrt{X}$ are shown
at different times. Outgoing and incoming waves are formed,
the incoming wave eventually bounces from the origin.}
  \label{fig000}
  \end{center}
\end{figure}
\begin{figure}[ht!]
  \begin{center}
 \epsfig{file=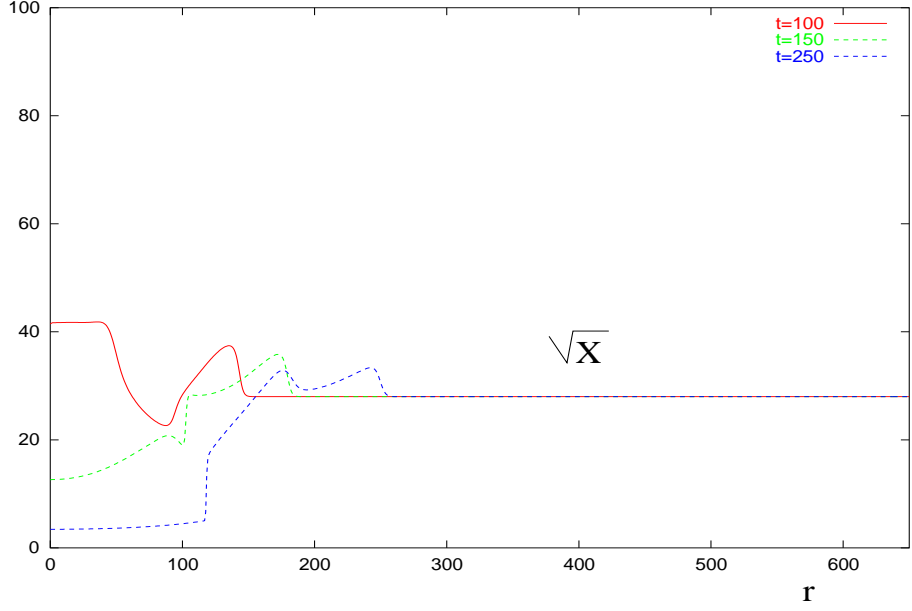,height=8cm,width=12cm} 
 \caption{The second outgoing wave forms a kink.
Note that the value of $X$ near the origin
temporarily becomes smaller than
in the original background.} 
  \label{fig100}
  \end{center}
\end{figure}
\begin{figure}[ht!]
  \begin{center}
 \epsfig{file=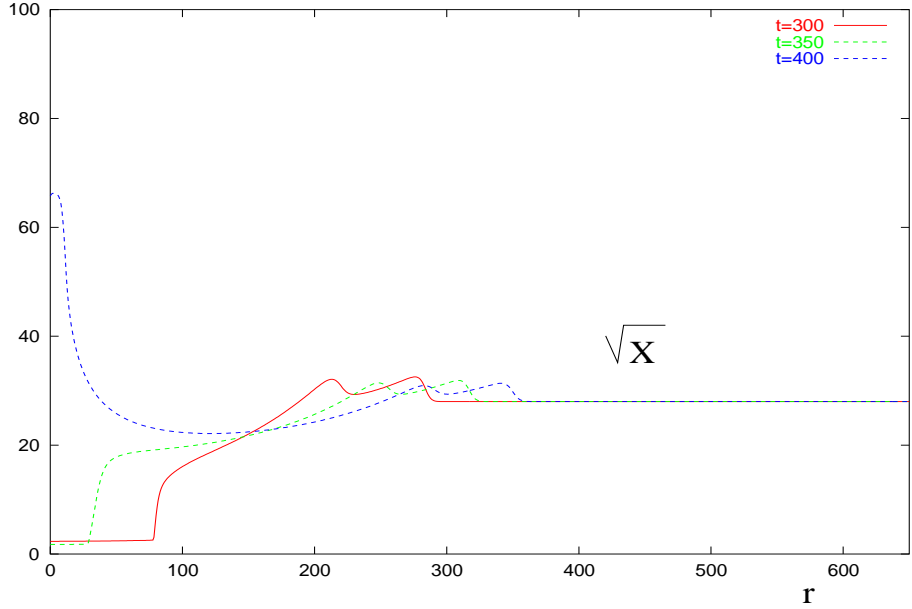,height=8cm,width=12cm} 
 \caption{The second kink moves back and hits the origin.}
  \label{fig300}
  \end{center}
\end{figure}
 \begin{figure}[ht!]
 \begin{center}
 \epsfig{file=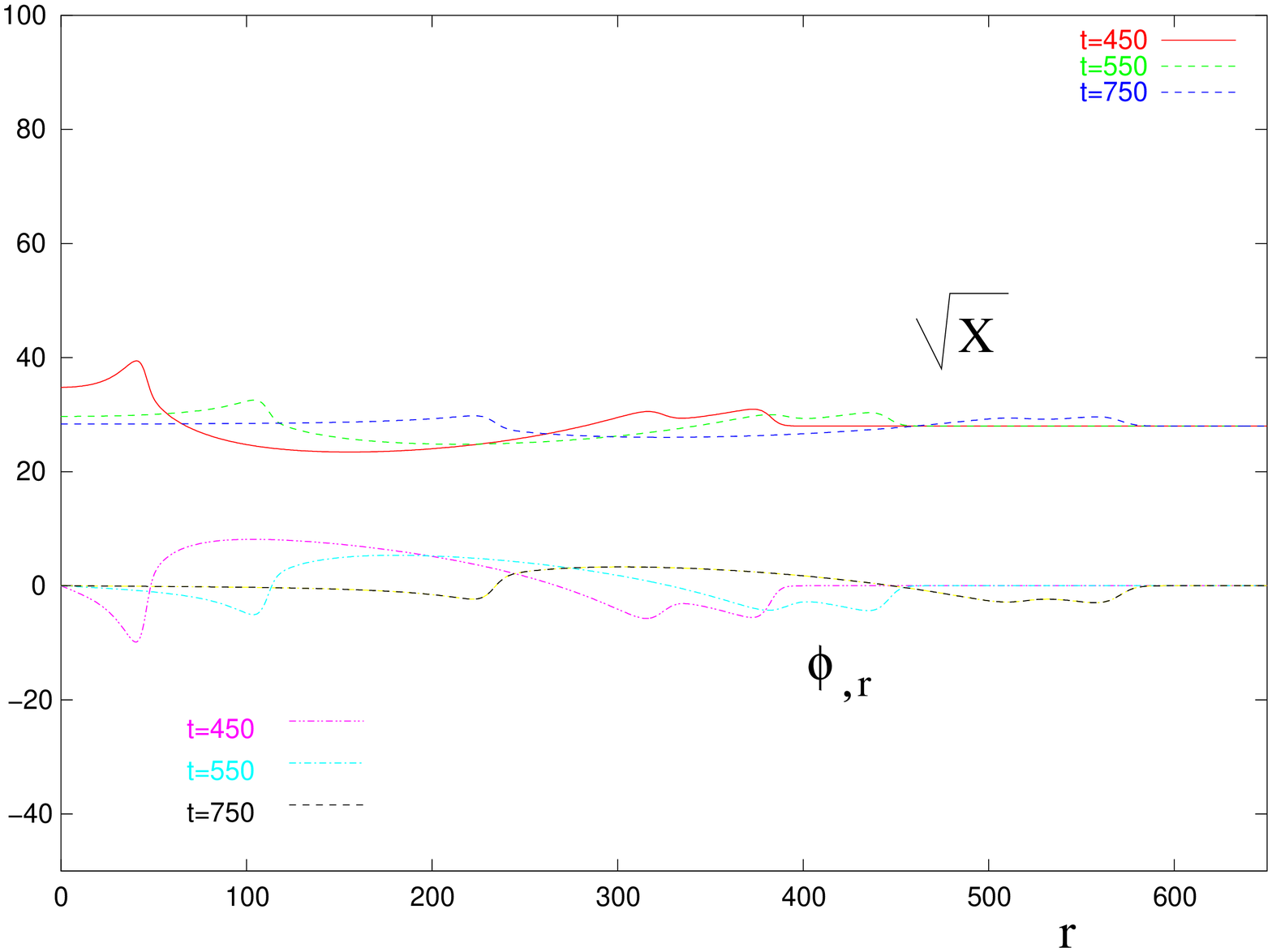,height=8cm,width=12cm} 
 \caption{The third outgoing wave is formed;
the background behind the outgoing waves settles down to its
original value.}
  \label{fig450}
  \end{center}
\end{figure}

In numerical simulations, we observed a fairly complex behavior of
the system even in the $O(3)$-symmetric case. As a reservation, we did
not incorporate the region where $X<c_*^2$ into our numerical study of
the long-time properties of the solutions. The reason is that the
system is unstable for $X<c_*^2$, and this instability 
becomes intolerable in
numerical simulations too early. Instead, we considered initial data
such that $\vec{\nabla} \phi = 0$, $\dot{\phi} \geq c_*$. To mimic
possible effects of the boundary of the stable region 
(a surface at which $X=c_*^2$), we considered
various boundary conditions near the origin, namely
(i) smoothness at the origin, $\phi_{,r}=0 $ at $r=0$;
(ii) free boundary condition at some fixed $r=r_h$ (meant to be the
radius of the hole); 
(iii) $X \equiv \phi_{,t}^2 - \phi_{,r}^2 = c_*^2$ at $r=r_h$.
The overall behavior of the solutions at $r \gg r_h$
was essentially independent of the choice of the boundary condition.

A typical sphericaly symmetric solution in the theory with the
potential (\ref{P}) is shown in figs.~\ref{fig000}~--~\ref{fig450}.
The initial data for this solution are such that $\phi_{,r} =0$ while
$\dot{\phi}$ has the bubble-like shape, as shown in fig.~\ref{fig000}.
The initial stages of the evolution, shown in 
figs.~\ref{fig000}~--~\ref{fig300}, are indeed quite complex, but at
late times the solution is simple: there is a sequence of outgoing
waves with the backgrond behind them equal to the background in front,
see fig.~\ref{fig450}. The amplitudes of the outgoing waves indeed
decrease in time, so the system settles down back to the homogeneous
configuration $\phi_{,r} =0$, $\dot{\phi} = c$.

Thus, the outcome of the entire process discussed in this paper
is a hole in ghost condensate plus a few outgoing
``ghost waves'' (in quantum theory the latter are
particles of the ghost field $\phi$).
Without knowing the UV-complete theory one cannot tell much about the
properties of the negative energy
holes. Our analysis  implies
that these holes do not expand, and thus remain of microscopic
size.
Although in the theory with the action (\ref{1**}) per se
the field inside the hole is unstable, it is conceivable that the state of
the hole gets stabilized due to UV effects. It is of interest to
further investigate the
phenomenology of ghost condensate with holes.
 
The authors are indebted to N.~Arkani-Hamed, F.~Bezrukov,
S.~Dubovsky, D.~Gorbunov, A.~Gorsky, D.~Levkov,
M.~Libanov, M.~Luty and S.~Sibiryakov for stimulating discussions.
This work was supported in part by RFBR grant 02-02-17398,
CRDF award RP-1-2364-MO-02 and US-DOE grant DE-FG02-91ER40676.
D.K. acknowledges the support by Dynasty Foundation.

\end{document}